# Tuning electrical properties of silicon dioxide through intrinsic nano-patterns

*Subimal Majee,[1][*] Devesh Barshilia,[1] Debashree Banerjee,[1] Sanjeev Kumar,[1] Prabhash Mishra[2] and Jamil Akhtar[1]*

[1]CSIR-Central Electronics Engineering Research Institute (CEERI), Pilani 333031, Rajasthan, India
[2]Nano-Science Center, Jamia Millia Islamia, New Delhi, 110025, India

**Corresponding Author:** *S. Majee (subimal.majee@polytechnique.edu)

ABSTRACT: The inherent network of nanopores and voids in silicon dioxide ($SiO_2$) is generally undesirable for aspects of film quality, electrical insulation and dielectric performance. However, if we view these pores as natural nano-patterns embedded in a dielectric matrix then that opens up new vistas for exploration. The nano-pattern platform can be used to tailor electrical, optical, magnetic and mechanical properties of the carrier film. In this article we report the tunable electrical properties of thermal $SiO_2$ thin-film achieved through utilization of the metal-nanopore network where the pores are filled with metallic Titanium (Ti). Without any intentional chemical doping, we have shown that the electrical resistivity of the oxide film can be controlled through physical filling up of the intrinsic oxide nanopores with Ti. The electrical resistance of the composite film remains constant even after complete removal of the metal from the film surface except the pores. Careful morphological, electrical and structural analyses are carried out to establish that the presence of Ti in the nanopores play a crucial role in the observed conductive nature of the nanoporous film.

1. INTRODUCTION

Nanopores and nano-voids inside thin films are usually undesirable due to the poor film quality, electrical insulation and dielectric performance. However, in recent decades materials with nano-pores have a wide spectrum of applications, like, fuel cells,[1] photovoltaic cells,[2] Lithium ion batteries,[3] supercapacitors,[4] catalysis,[5] gas purification[6] and sensors[7]. In most of these applications, a precise size distribution of the nano-pores is required, which is generally fabricated by some extrinsic methods. Owing to their nanometer dimensions, such nanopore networks have the potential to provide tailored optical, electrical, chemical, mechanical and magnetic properties in the carrier material.[8-15] In the oxide materials, like, Silicon dioxide ($SiO_2$), there exist inherent networks of nanopores which are process independent. $SiO_2$ is one of the most widely researched materials in silicon technology for the last few decades. Nanoporous $SiO_2$ thin film has been recently attractive due to its different applications, such as, low dielectric coefficient materials in electronics and in biological sensors.[16] It has also been shown that the piezoelectric effect of such nanoporous films improves the sensitivity of the biosensors due to their higher surface areas at the nanoporous structures.[17, 18]

In this study, we have utilized the intrinsic nanopores inside $SiO_2$ thin films which can be considered to be inherent nano-patterns, as a platform to tune the $SiO_2$ thin film properties. We showed here the tunable electrical properties of the $SiO_2$ thin film through careful and simple implementation of the nano-patterns. To the best of our knowledge, this simple process to implement the intrinsic nano-patterns to tune the properties of the oxides has not been reported elsewhere before. The intrinsic nano-patterns inside the thermally grown $SiO_2$ thin films are physically filled in by Titanium (Ti) metal. Electrical conduction in the oxide layer is demonstrated even after complete removal of the metal from the oxide surface except the nanopores. Without any intentional expensive chemical doping process, we are able to fabricate tunable conductive oxide layers which have the potential to be used in the above mentioned applications.

2. RESULTS AND DISCUSSIONS

The as-grown $SiO_2$ (initial thickness 200 nm) is coated with Ti through e-beam evaporation, which is etched away gradually from the oxide surface. The variation of the average surface roughness of the film before and after metal etching is shown in **Figure 1(a)**. For reference, metal deposited on bare Si wafer is also shown. Three distinct regions are observed for both the samples. We observe a monotonic increase of the average surface roughness (region I) up to 10 s for the Ti/$SiO_2$/Si sample and up to 7 s for the Ti/Si sample, respectively. This effect is called roughening effect, which is due to a gradual dissolution of the outer Ti layer. At the end of this region, all the metal from the top surfaces have been removed and the maximum roughness values achieved are $1.25 \pm 0.05$ nm for the Ti/$SiO_2$/Si sample and $0.96 \pm 0.04$ nm for the Ti/Si sample, respectively. The surface roughness decreases monotonically afterwards in region II and becomes stable in region III for the Ti/$SiO_2$/Si sample. However, a rapid decrease of the surface roughness is observed in region II for the Ti/Si sample which becomes stable at the initial Si wafer roughness value ($0.12 \pm 0.03$ nm). AFM morphological images of the Ti/$SiO_2$/Si sample before metal etching and after partial etching (<10 s) have also been shown in **Figure 1(a)** at their corresponding data points.

The maximum pore depth is estimated from the topographical analysis and a linear enhancement of the pore depth is observed after 10 s of etching duration, as shown in **Figure 1(b)**. From the AFM topographical image at 10 s, discrete metal particles with average size of ~30 nm are observed. After a critical etching time ($t_c$) of ~13 s the initial nano-pattern of the oxide layer becomes visible. However, we estimate that at this stage the nano-pores are partially filled with the metal since the pore depth (7.2 nm) at this stage is lesser compared to the initial pore depth of as-grown oxide sample. With increasing etching time, the metal is monotonically removed from the nano-pores. The final pore depth of ~ 9 nm is similar to the initial pore depth of the oxide layer suggesting that at this stage all the metal from the nano-pattern is been etched out. From the AFM morphological images in **Figure 1(b)** the nano-pattern of the film is clearly visible. We observe nano-patterns (at 16 s) inside the layer with nano-pore dimensions of around 30 nm and pore-to-pore distance around 50 nm in average, which is comparable with the initial pore size and pore-to-pore distance in as grown $SiO_2$ thin film.

**Figure 1(b) right side** shows the variation of the porosity of the sample with gradual removal of metal from the surface. The porosity ($P$) is calculated from the analysis of the morphological images and assuming the nanopores to be cylindrical in nature.[19] With gradual removal of the metal from the surface and from the pores after the critical etching time (10 s), the porosity of the sample increases linearly reaching a maximum value of around 3%. It is to be noted that, the maximum porosity obtained after 16 s of etching duration is similar to the initial porosity of the as-grown oxide film and thus confirming indirectly that at this stage all the metal have been washed away from the sample leaving the initial nano-pattern.

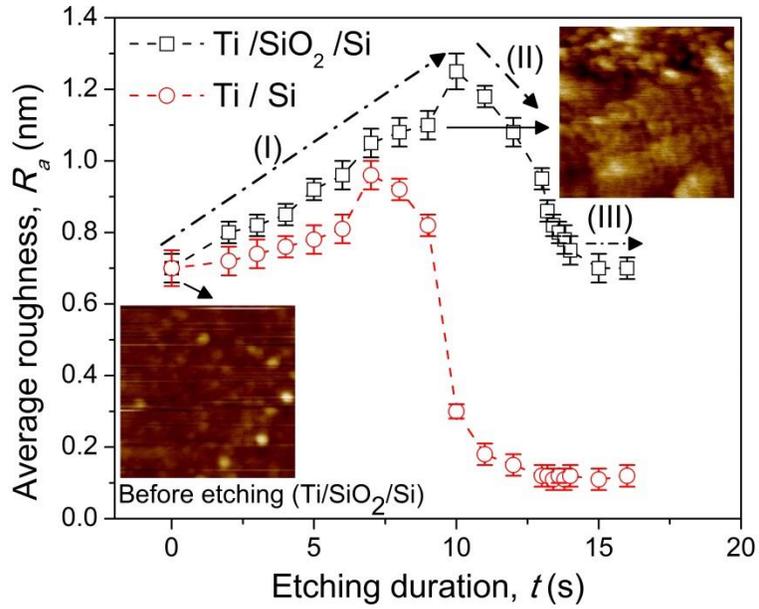

(a)

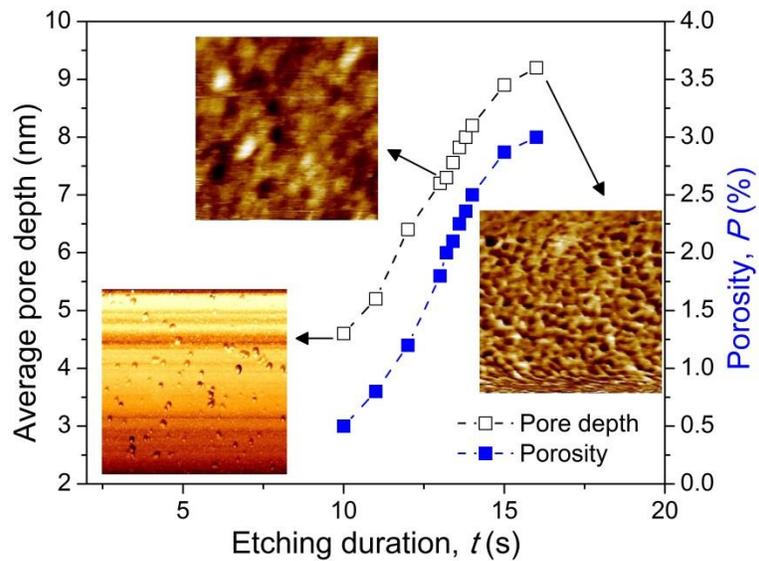

(b)

**Figure 1**: (a) Variation of average surface roughness with etching duration. Ti on bare Si wafer is shown for reference. (b) Change of the average pore depth and porosity with etching duration. AFM topographical images (500 nm × 500 nm) are shown at their corresponding data points.

**Figure 2** shows the FE-SEM images for the Ti/SiO$_2$/Si sample after different etching durations. It is clearly observed that the metal is gradually removed from the sample surface with the increase of the etching duration. When the etching duration is less than 10 s, we observe layers of metal on the top surface. At 10 s of etching time, discrete distributions of clusters of metal particles are visible on the sample surface, which remains intact even at 12 s of etching time. It is interesting to note that the same discrete metal clusters with average size of ~30 nm are also observed through AFM measurements as mentioned above. At the final stage (after 14 s) nano-patterns are visible resembling with the nano-pattern size distribution as observed through AFM measurements for the as grown oxide thin film. This observation indirectly confirms the trapping of metal particles inside the nanopores only whereas the metal is completely washed away from the other areas of the top surface. Due to the presence of nano-pores inside the oxide layer, the surface energy is different at the pore edges compared to the flat oxide surface.[20] The nano-pores are thus more "active" compared to the 'dead' flat surfaces and there is a natural tendency of the metal atoms to accumulate at the higher surface energy sites as compared to the flat surfaces. The presence of clusters of metal particles is possibly due to the cohesive attraction between the outside metal particles and the particle lying inside the nanoporous structure. Because of their strong interaction, the etching rates at those sites are different from the outer surface, creating discrete clusters like structures over the nano-pores. When the sample is further etched, at around 14 s, the initial nano-pattern of the oxide layer becomes visible.

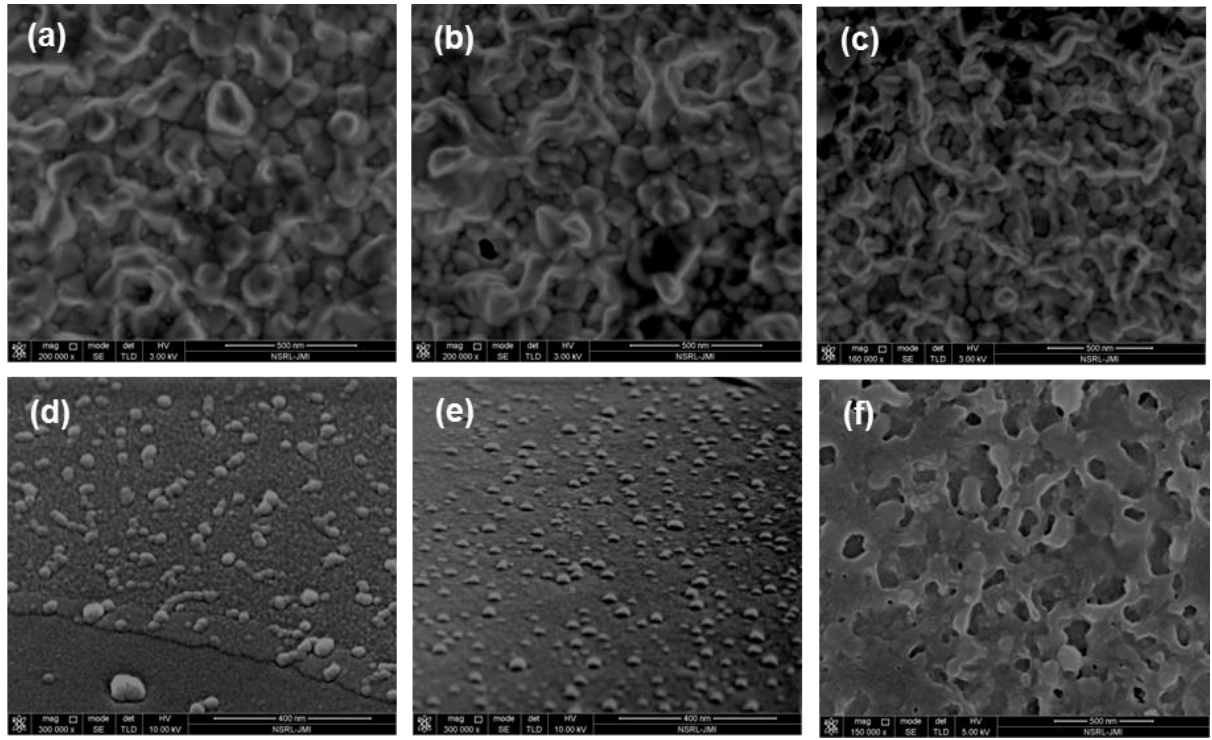

**Figure 2**: FESEM images for the Ti/SiO$_2$/Si sample after different etching durations: (a) 2 s, (b) 5 s, (c) 8 s, (d) 10 s, (e) 12 s and (f) 14 s.

**Figure 3(a)** shows the electrical performance of the samples with the variation of the metal etching duration. The sample based on bare Si wafer is shown as a reference. The initial highly conductive films become gradually non-conductive with the increase of the etching durations. An identical trend is observed for the electrical resistance data as compared to the surface roughness measurements and pore depth analysis. The initial electrical resistance of the sample on bare Si wafer (Ti/Si) remains constant until an etching duration of only 7 s, whereas the electrical resistance remains fixed up to 11 s of etching time for sample with oxide layer (Ti/SiO$_2$/Si). A zoom in image of the initial variation of the electrical resistance values for the sample on the oxide layer is shown in **Figure 3(b)**, where a monotonic increase of the resistance can be observed. This is due to the fact that with increase of the etching duration, the metal layer becomes thinner and the resistance increases accordingly. Two different slopes are observed for the oxide sample, indicating two different phases of the whole process. The first slope, where the resistance increases linearly with the etching time, is due to time dependent removal of the metal from the surface of the samples. We conclude that the longer persistence of the electrical conductivity for the oxide sample is due to the metal-nanopore network, where the etching rate is different due to the higher surface energy at the pore edges. We observe an exponential increase of the electrical resistance for the oxide sample afterwards, confirming indirectly the presence of metal inside the pores (not on the surface). Due to non-existence of nano-pattern in Ti/Si sample, the Ti is etched away rapidly and the electrical resistance immediately increases only after 7 s. At the final stage, the sample based on bare Si wafer shows the initial resistance of the high-resistive Si wafer and the sample based on oxide layer shows the insulating nature of the oxide. From these observations it can be concluded that the nano-pores inside the oxide layer play a crucial role in tuning the electrical resistance values by trapping the metal particles inside their pits. Further investigations are carried out in order to confirm the presence of metal inside the nano-pores which is described below.

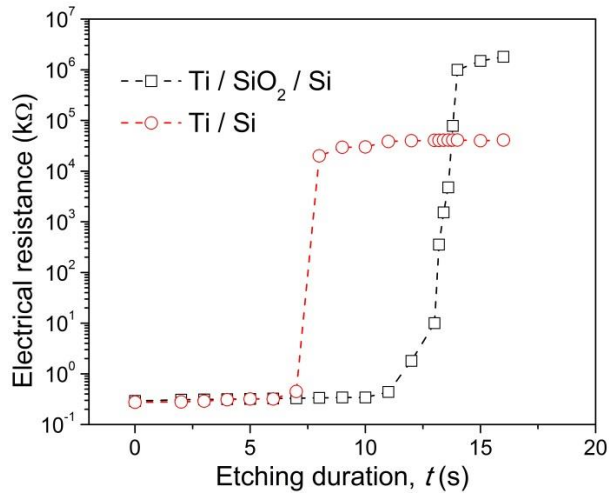

(a)

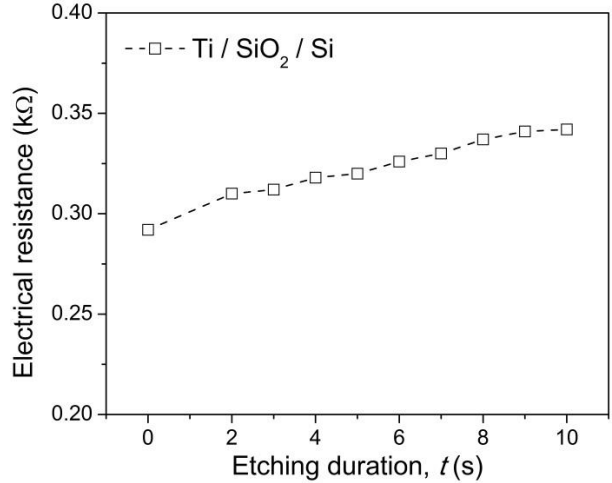

(b)

**Figure 3**: (a) Variation of electrical resistance with Ti etching duration. Ti on Si is shown as a reference. Each data points are averaged over 10 different measurements. (b) Zoom in image of the Ti etching up to 10 s.

**Figure 4** shows the Raman spectra for the Ti / SiO$_2$ / Si sample after different etching durations. Symmetric Ti-O-Si stretching mode is observed at around 800 cm$^{-1}$, which indicates the presence of Ti over the sample.[21] With gradual removal of the metal, we observe that the Ti-O-Si peak remains intact up to 10 s of etching duration. It is to be noted that we observe no indication of Ti after 6 s of etching from the Ti / Si sample. The presence of Ti up to 10 s of etching for the oxide sample is likely due to the mixed metal-nanopore network which contributes to the Ti-O-Si stretching mode. With further removal of the metal from the structure, the Si-O peak appears at around 1000 cm$^{-1}$ and no further sign of metal over it. At this stage, the metal is completely removed and the insulating oxide layer appears which supports the electrical characterization data in **Figure 3(a)**.

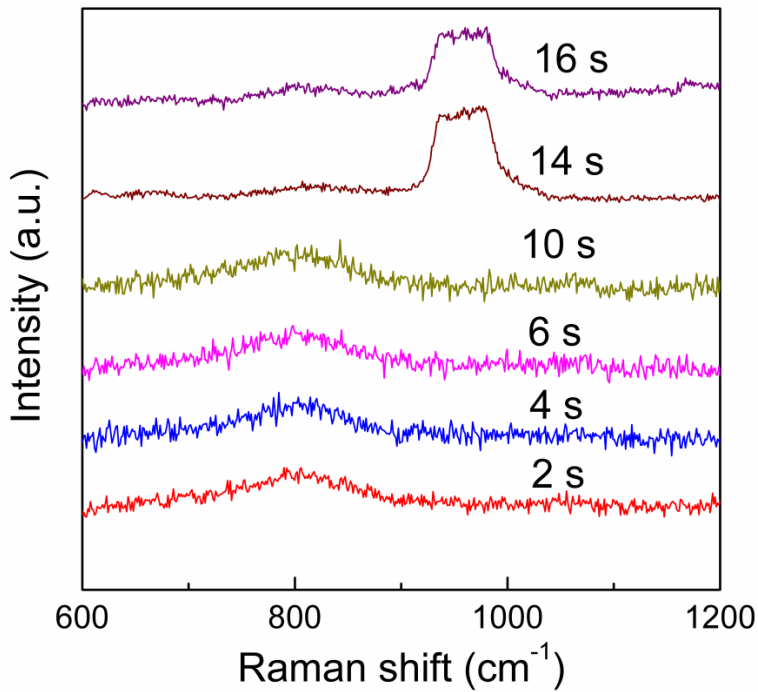

**Figure 4**: Raman shifts for the sample with different Ti etching durations averaged over 5 different measurements.

Energy-dispersive X-ray spectroscopy (EDS) analysis has been carried out with the Ti/SiO$_2$/Si sample after different etching durations. **Figure 5** shows the EDS elemental mapping of the Ti/SiO$_2$/Si sample after 10 s of etching duration. Details of the EDS measurements are provided in the Supplementary information. As observed from the EDS analysis (**Figure S1**), with the gradual increase of the etching duration, the atomic percentage of Ti diminishes steadily. After 8 s of etching duration, presence of discrete dots of Ti metal is visible; this gradually vanishes with further etching. There is no indication of Ti after an etching duration of 16 s. The EDS analysis results correlated well with the morphological and electrical characterizations as mentioned above.

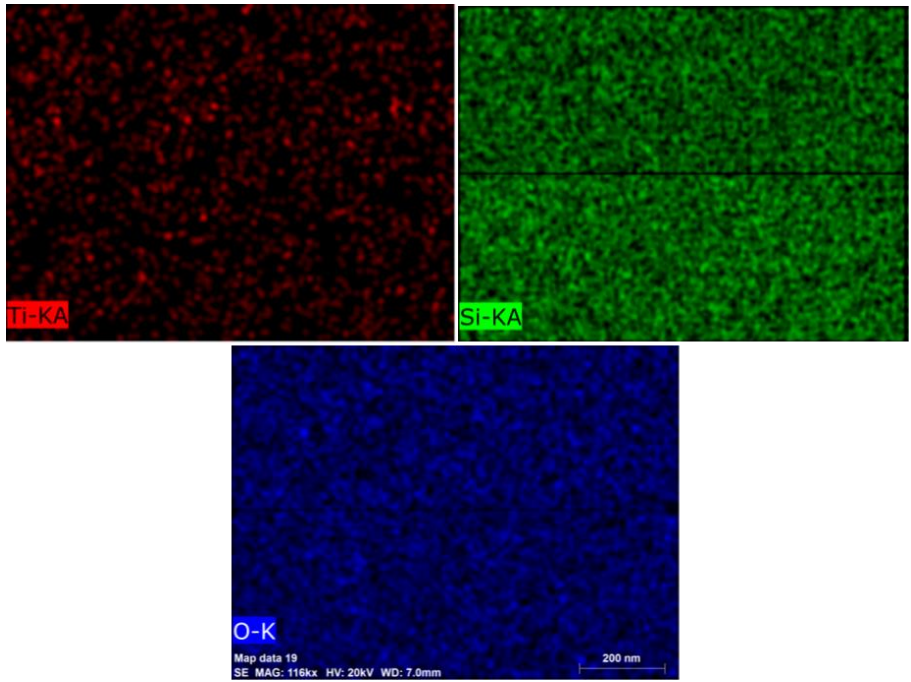

**Figure 5**: EDS elemental analysis of the Ti/SiO$_2$/Si sample after 10 s of etching duration (scale bar 200 nm).

3. CONCLUSIONS

In summary, a simple method of fabricating conductive oxide layer is shown in this work. Without any intentional chemical doping process, we are able to tune the conductive nature of the oxide film through simple metal filling inside the inherent nano-pattern. We have utilized the intrinsic nano-pattern as a platform to tailor the electrical properties of the carrier oxide film. Morphological, structural and electrical characterizations are carried out in order to understand the role of the nano-patterns in tuning the electrical properties of the metal filled nanoporous oxide layer. This study indicates that each of the undesirable nanopores inside the oxide thin films has the potential to be an active device if sufficiently tuned.

METHODS

*1) Oxide film growth and metal filling inside nanopores:*

In this study the oxide film is grown through the standard thermal oxidation process on a cleaned high resistive silicon wafer. The initial oxide thickness is 200 nm. Titanium (200 Å) and gold (2000 Å) are deposited onto the oxide in a high vacuum ($10^{-7}$ Torr) e-beam evaporation system, consecutively. Gold is deposited on top of Ti in order to avoid immediate oxidation of Ti and removed from the surface using standard gold etchant and Ti is etched carefully with different time variations using diluted HF solution at room temperature where the concentration of the solution remains constant.

*2) Electrical characterizations:*

For the electrical measurements, samples are diced into 1.5 cm square sizes. Electrical contacts (2 mm width) are fabricated at the two opposite sides of the samples using silver paste and then dried. Electrical measurements are carried out using a Keithley probe station.

*3) Other characterizations:*

Morphological and elemental analyses are carried out by tapping mode AFM, Field Emission Scanning Electron Microscope (FE-SEM), Energy-dispersive X-ray spectroscopy (EDS), Raman measurements.

**Author Contributions**

The manuscript was written through contributions of all authors. All authors have given approval to the final version of the manuscript. SM, DB (Devesh Barshilia) and JA contributed in planning, designing and implementation of the study; SM contributed in writing the manuscript; PM contributed in FE-SEM and EDS analysis; SK and DB (Debashree Banerjee) contributed in useful discussions.

**Competing Financial Interests statement**

The authors acknowledge that there is no Competing Financial Interests.

ACKNOWLEDGMENT

Prof. Santanu Chaudhury, Director CEERI-Pilani is thankfully acknowledged for his support and encouragement. The authors are thankful to Mr. G. S. Negi for thermal oxide growth.

REFERENCES

[1] K. L. Yeung and W. Han, Catalysis Today, 236, 182 (2014).

[2] G. Shi and E. Kioupakis, ACS Photonics, 2, 208 (2015).

[3] M. Ge, X. Fang, J. Rong and C. Zhou, Nanotechnology, 24, 422001 (2013).

[4] Y. Ding and Z. Zhang, Nanoporous Metals for Advanced Energy Technologies, (Springer, 2016).

[5] G. Sneddon, A. Greenaway and H. H. P. Yiu, Adv. Energy Mater. 4, 1301873 (2014).

[6] D. P. Broom and K. M. Thomas, MRS Bulletin, 38, 412 (2013).

[7] K. Jiao, K. T. Flynn and P. Kohli, "Synthesis, Characterization, and Applications of Nanoporous Materials for Sensing and Separation" in Handbook of Nanoparticles, edited by M. Aliofkhazraei (Springer, 2016).

[8] G. Moorthy and K. Daneshvar, Journal of Applied Physics, 111, 124503 (2012).

[9] L. M. Manocha, "Composites with nanomaterials," in Functional Nanomaterials, edited by E. Rosenberg and K. E. Geckeler (American Scientific, California, 2006).

[10] R. Davies, G. A. Schurr, P. Meenan, R. D. Nelson, H. E. Bergna, C. A. S. Brevett, and R. H. Goldbaum, Adv. Mater. 10, 1264 (1998).

[11] H. Ow, D. R. Larson, M. Srivastava, B. A. Baird, W. W. Webb, and U. Wiesner, Nano Lett. 5, 113 (2005).

[12] N. Sounderya and Y. Zhang, Recent Pat. Biomed. Eng. 1, 34 (2008).

[13] C. Sanchez, B. Julian, P. Belleville, and M. Popall, J. Mater. Chem. 15, 3559 (2005).

[14] S. M. Kang, K. Lee, D. J. Kim, and I. S. Choi, Nanotechnology 17, 4719 (2006).

[15] L. M. Liz-Marzan, M. Giersig, and P. Mulvaney, Langmuir 12, 4329 (1996).

[16] F. Chen and A. H. Kitai, Thin Solid Films, 517, 622 (2008).

[17] A.F. Collings and F. Caruso, Rep. Prog. Phys. 60, 1397 (1997).

[18] J.F. Diaz and K.J. Balkus, J. Mol. Catal. B: Enzymatic 2 (1996).

[19] F. Alfeel, F. Awad, I Alghoraibi and F. Qamar, Journal of Materials Science and Engineering A, 2 (9), 579 (2012).

[20] S. Majee, PhD thesis. Ecole polytechnique, France, Development of efficient permeation barriers based on hot-wire CVD grown silicon nitride multilayers for organic devices deposited on flexible substrates, 2014 https://pastel.archives-ouvertes.fr/pastel-01067937/document.

[21] G. Ricchiardi, A. Damin, S. Bordiga, C. Lamberti, G. Spano, F. Rivetti and A. Zecchina, J. Am. Chem. Soc., 123, 11409 (2001).

# Supplementary information

# Tuning electrical properties of silicon dioxide through intrinsic nano-patterns

*Subimal Majee,[1*] Devesh Barshilia,[1] Debashree Banerjee,[1] Sanjeev Kumar,[1] Prabhash Mishra[2] and Jamil Akhtar[1]*

[1]CSIR-Central Electronics Engineering Research Institute (CEERI), Pilani 333031, Rajasthan, India
[2]Nano-Science Center, Jamia Millia Islamia, New Delhi, 110025, India

**Corresponding Author:** *S. Majee (subimal.majee@polytechnique.edu)

*EDS measurements*

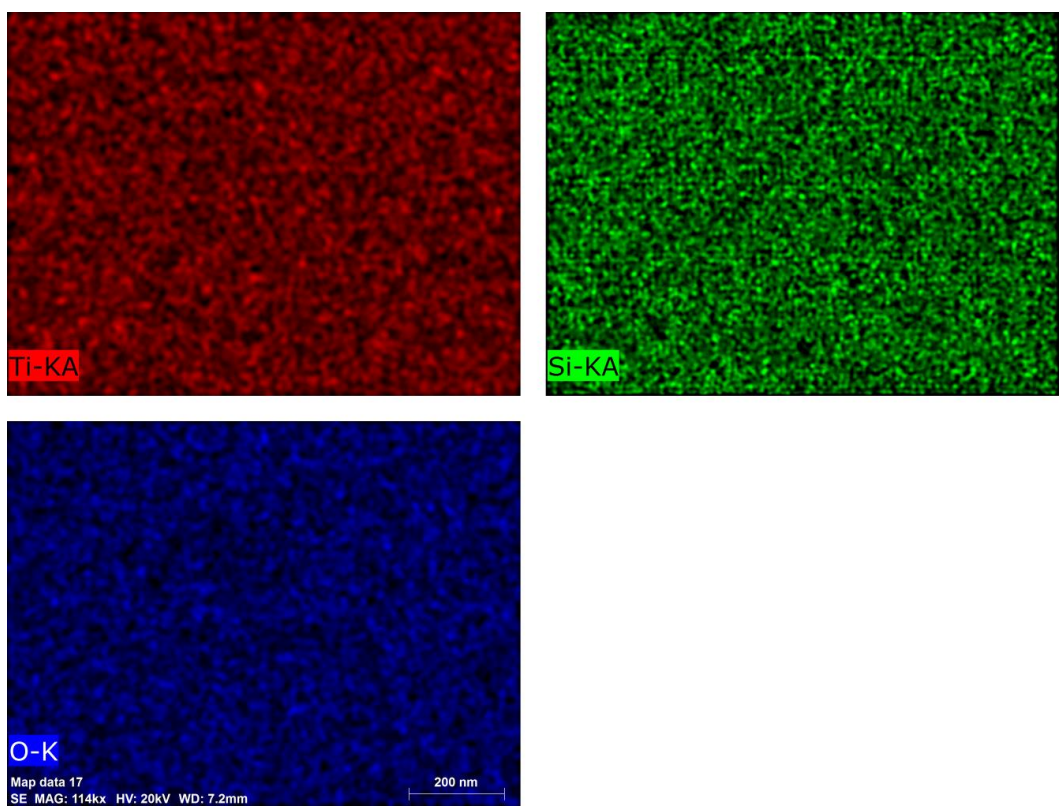

(a)

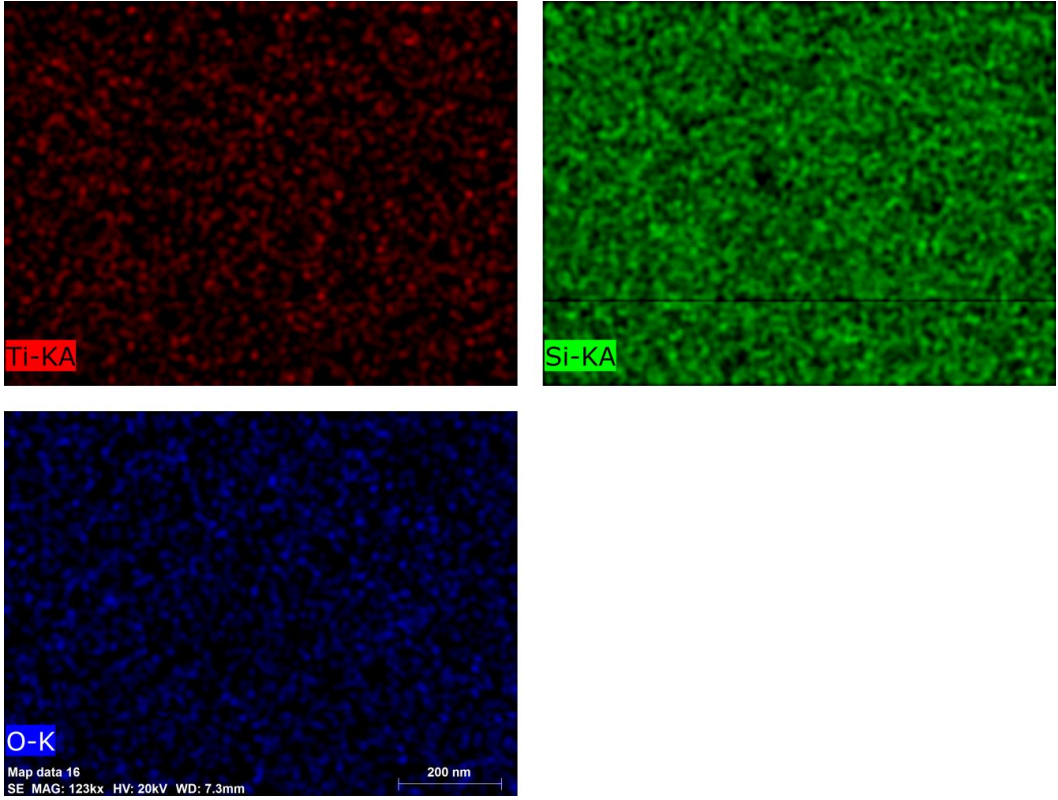

(b)

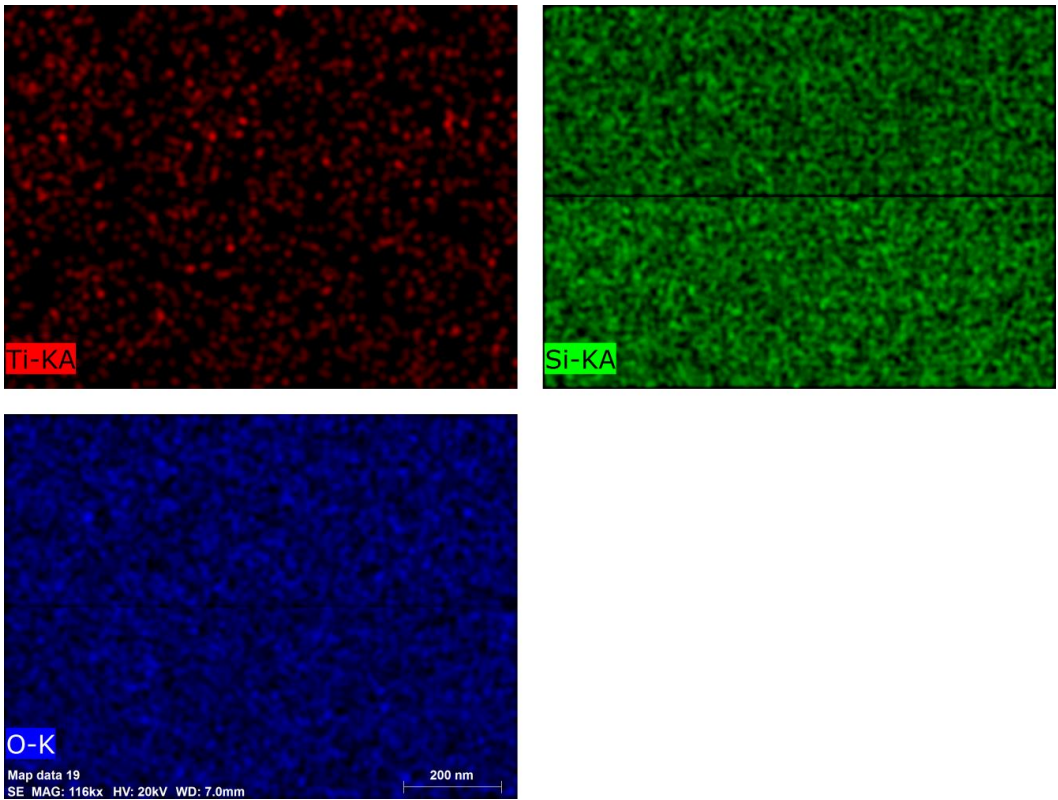

(c)

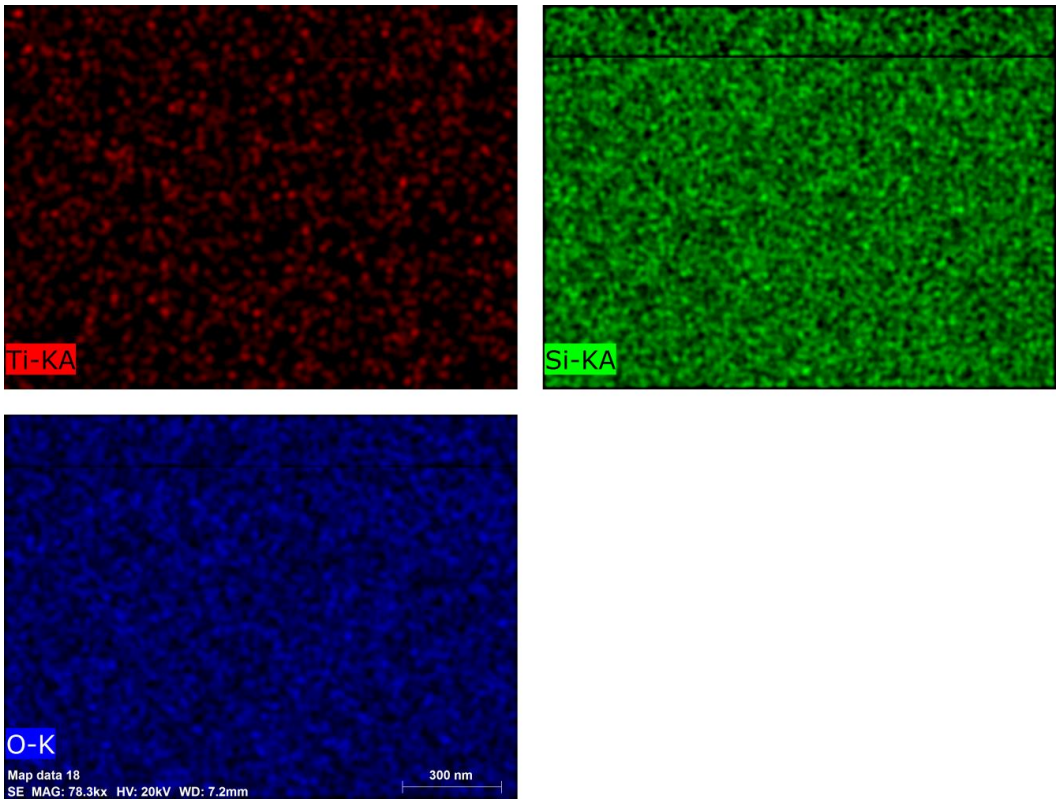

(d)

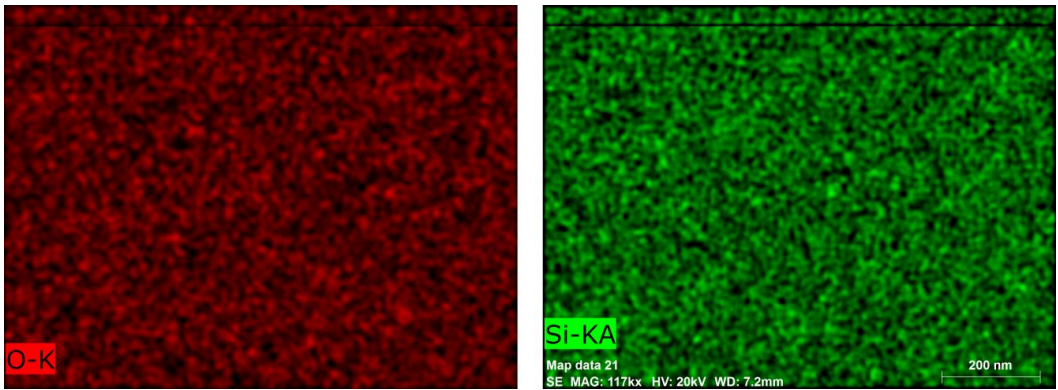

(e)

**Figure S1**: EDS elemental results for the Ti/SiO$_2$/Si sample after different Ti etching durations: (a) 2 s, (b) 8 s, (c) 10 s, (d) 12 s and (e) 16 s.